\documentclass[
  reprint,
 superscriptaddress,
 amsmath,amssymb,
 aps,
pra,
]{revtex4-2}
\usepackage{orcidlink}
\usepackage{graphicx}
\usepackage{physics}
\usepackage{xspace}
\usepackage{amsmath}
\usepackage{hyperref}
\usepackage{color}
\usepackage[normalem]{ulem}

\newcommand{\im}{{\rm i}}
\newcommand{\ps}{phase space\xspace}
\newcommand{\cterm}{identity-of-outcome\xspace}
\newcommand{\VEC}[1]{{\mbox{\boldmath${#1}$}}}

\newcommand{\NO}[1]{{\color{green}}}

\makeatletter
\newsavebox{\@brx}
\newcommand{\llangle}[1][]{\savebox{\@brx}{\(\m@th{#1\langle}\)}%
  \mathopen{\copy\@brx\kern-0.5\wd\@brx\usebox{\@brx}}}
\newcommand{\rrangle}[1][]{\savebox{\@brx}{\(\m@th{#1\rangle}\)}%
  \mathclose{\copy\@brx\kern-0.5\wd\@brx\usebox{\@brx}}}
\makeatother

\usepackage{forloop,ifthen} 

\usepackage{xifthen}
\newcommand{\refAppendix}[6]{#1
  \ifthenelse{\isempty{#2}}%
    {}
    {\protect\cite{#2}}
    #3\protect\ref{#4}#5#6\xspace
}

\begin{document}

\title{Adding or Subtracting a single Photon is the same for Pure Squeezed Vacuum States}

\author{Ole Steuernagel\orcidlink{0000-0001-6089-7022}}
\email{Ole.Steuernagel@gmail.com}
\affiliation{Institute of Photonics Technologies, National Tsing Hua University, Hsinchu 30013, Taiwan}
\affiliation{Tulane University, New Orleans, LA 70118, USA}

\author{Ray-Kuang Lee\orcidlink{0000-0002-7171-7274}}
\email{rklee@ee.nthu.edu.tw}
\affiliation{Institute of Photonics Technologies, National Tsing Hua University, Hsinchu 30013, Taiwan}
\affiliation{Department of Physics, National Tsing Hua University, Hsinchu 30013, Taiwan}
\affiliation{Center for Theory and Computation, National Tsing Hua University, Hsinchu 30013, Taiwan}
\affiliation{Center for Quantum Science and Technology, Hsinchu, 30013, Taiwan}
 
\date{\today}
\begin{abstract}
  The addition of a single photon to a light field can lead to exactly the same \emph{outcome} as
  the subtraction of a single photon. We prove that this \cterm is true for pure squeezed vacuum
  states of light, and in some sense only for those. We show that mixed states can show this \cterm
  for addition or subtraction of a photon if they are generated from incoherent sums of pure
  squeezed vacuum states with the same squeezing. We point out that our results give a
  reinterpretation to the fact that pure squeezed vacuum states, with squeezing~$e^{-z}$, are formally
  annihilated by Bogoliubov-transformed annihilation
  operators:~$\hat a_z =  \hat a \cosh(z) - \hat a^\dagger \sinh(z) $.
\end{abstract}

\maketitle 

\section{Introduction}

Adding a single photon to a light field, or subtracting a single photon are different processes.  So
different that it seems the question as to whether a quantum state can exist that gives the same
 \emph{outcome} after a photon is subtracted from it, or a photon is added to it, has not been
investigated, or at least not been answered in the affirmative~\cite{Chung_MPLA14,Lu__CPL99,Lvovsky__A20}.

It might appear surprising that adding a photon or subtracting one can give the same outcome, for
correctly chosen input states, seemingly viola\-ting energy conservation. Yet, here, in
Sect.~\ref{sec:Schroedinger-Formulation}, using Schr\"odinger wave functions, we prove that this
surprising fact holds true for all \emph{pure squeezed vacuum} states.  Then, in
Sect.~\ref{sec:Wign-Distr-Formulation}, we introduce Wigner's \ps formulation and apply it, in
Sect.~\ref{sec:a_adag_ps}, to consider the case of mixed gaussian states as input states and show
that this \cterm is only true for pure squeezed vacuum input states.  In Sect.~\ref{sec:PureFock},
we explicitly show that this \cterm also holds when using the Fock representation. We then show, in
Sect.~\ref{sect:ImpureSqueezedVacua}, that only incoherent sums of pure squeezed vacuum states with
the same amount of squeezing, as input states, can give \cterm.  We conclude with some observations
about the technological rele\-vance and meaning of our results.

\section{Using Wave Functions~\label{sec:Schroedinger-Formulation}}

We consider a bosonic single mode system, specifi\-cally a harmonic oscillator whose creation
operator is~$\hat a^\dag(\hat x, \hat p)$ $=$
$\sqrt{\frac{m \omega }{2 \hbar }} (\hat x-\frac{ \im \hat p}{m \omega })$. Setting $m$=1 and
$\omega$=1, $\hat a^\dag$ can describe the excitation of an optical mode~\cite{Cahill_PR69a} by the
addition of a single photon, for brevity we often also set $\hbar$=1.

Since the momentum operator has the form $\hat p = \frac{\hbar }{\im} \frac{\partial}{\partial x}$,
the corresponding annihilation operator, describing the removal of a single photon, is
$\hat a = (\hat x+ \frac{\partial}{\partial x})/\sqrt{2}$.

It is well known that the removal of a single photon has the optical vacuum state as its trivial
eigensolution, in Fock language: $\hat a | 0 \rangle = 0$. Addition of a photon has no eigenstate
(formally, it leads to a non-normalizable gaussian state~\cite{Miller_PR66}).

Here, instead of finding eigenstates, we identify input states for which addition or subtraction of
a photon leads to \cterm, we ask:

Do quantum states, $\psi$, exist, which, after addition,~$\psi_+$, or subtraction,~$\psi_-$, of
a photon,  give the same outcome state:~$\psi_+ = \psi_-$? \\
Formally, for such pure input states~$\psi(x)$, this means we have to equate the operations
\begin{eqnarray}
  \label{eq:AnniCrea_psi_1}
  \hat a^\dag  \cdot \psi(x) = r \;  \hat a \cdot \psi(x) 
\end{eqnarray}
for the resulting \emph{renormalized} outcome states. Namely, we have to confirm that the \cterm
condition~(\ref{eq:AnniCrea_psi_1}) is fulfilled whilst
\begin{eqnarray}
  \label{eq:AnniCrea_r_Norm_condition}
  r = \sqrt{|\!|\hat a^\dag \psi |\!|_2 / |\!|\hat a \psi |\!|_2} \; ,
\end{eqnarray}
of Eq.~(\ref{eq:AnniCrea_psi_1}), forms the ratio of norms~(\ref{eq:AnniCrea_r_Norm_condition}),
where $ |\!| \psi |\!|_2 = \int_{-\infty}^{\infty} dx |\psi(x)|^2 $ is the regular $L_2$-norm.

Multiplying both sides with~$\sqrt{2}$ and separating terms, Eq.~(\ref{eq:AnniCrea_psi_1}) amounts
to 
\begin{eqnarray}
  \label{eq:AnniCrea_psi_2}
  (1 - r) \;  x \;  \psi(x) = ( 1 + r ) \frac{\partial}{\partial x}  \psi(x) \; .
\end{eqnarray}

This is obviously solved by gaussian wave functions of the form
  \begin{eqnarray}
\label{eq:pure_squeezed_psi}
  \psi(x, \sigma_x) = \exp [-x^2/(2 \sigma_x^2)]\left/\sqrt{\sigma_x \sqrt{ \pi}} \right. \; \; .
  \end{eqnarray}
In \ps, state~(\ref{eq:pure_squeezed_psi}) is aligned with the $x$-axis.

We emphasize that, for $\sigma_x = 1$, $ \psi(x, 1) = \psi_{|0\rangle}(x) $ is the vacuum state,
since $ |\psi_{|0\rangle}(x)|^2 $$=$ $ \exp [-x^2/(2 \Sigma_0^2)]/\sqrt{2 \pi \Sigma_0^2}$
correctly implies that $\Sigma_0^2 = \sigma_x^2/2 = \frac{1}{2} \frac{\hbar}{m \omega}$.

Checking whether $r$ represents the ratio of norms~(\ref{eq:AnniCrea_r_Norm_condition}), we find,
when using pure normalized squeezed vacuum states~(\ref{eq:pure_squeezed_psi}) as input states,
Eq.~(\ref{eq:AnniCrea_psi_2}) yields the conditions that $\sigma_x > 1$ and that
$ r(\sigma_x) = \sqrt{\frac{\left(\text{$\sigma_x
        $}^2+1\right)^2}{\left(\text{$\sigma_x $}^2-1\right)^2}} =
\frac{\left(\text{$\sigma_x $}^2+1\right)}{
  \left|\text{$\sigma_x $}^2-1\right|} $.

This is just what Eq.~(\ref{eq:AnniCrea_r_Norm_condition}) implies and establishes that
all pure squeezed states, anti-squeezed in position (for them $\sigma_x>1$), are states for which
addition of a photon or subtraction of a photon gives the same outcome. 

\subsection{Resolving the sign problem~\label{subsec:SignProbelm}}

This result for~$r(\sigma_x)$, just given, appears peculiarly limited since in quantum \ps position
and momentum essentially play equivalent roles~\cite{Cahill_PR69a} and it is therefore unclear why
only states anti-squeezed in position ($\sigma_x>1$), and thus squeezed in momentum should show this
\cterm. What about states squeezed in position ($0<\sigma_x<1$), and anti-squeezed in momentum?

Some consideration reveals that, in Eq.~(\ref{eq:AnniCrea_r_Norm_condition}), we dropped the negative
branch of the square root.  Substitution of $r$ with $-r$ in Eq.~(\ref{eq:AnniCrea_psi_1}) addresses
this oversight: with this sign convention all position-squeezed pure states ($0<\sigma_x<1$) fulfil
the desired \cterm as well.

We conclude that all pure squeezed vacuum input states, as long as $\sigma_x \neq 1$, yield an
\cterm, regardless of whether a photon is added or subtracted.

There are two good ways to circumvent this sign problem we have just identified:

For position-squeezed input states ($0<\sigma_x<1$) the outcome of the operation $\hat a \psi(x)$
switches the sign such that resulting wave functions have negative values for $x<0$. This is in
contravention of the sign convention of Hermite's polynomials of order~$n$ for the harmonic
oscillator, which are adorned by the coefficient $(-1)^n$, and this switching-problem can therefore
be `repaired' by multiplying with the sign-function sgn$(\sigma_x^2 - 1) \; \hat a \cdot \psi(x)$.

Instead, here we take the concise approach of taking the branches of the square
root~(\ref{eq:AnniCrea_r_Norm_condition}) such that the sign automatically switches by itself, we
use
\begin{eqnarray}
  \label{eq:AnniCrea_r_condition_1}
  r(\sigma_x) =  \frac{\left(\text{$\sigma_x $}^2+1\right)}{
  \left(\text{$\sigma_x $}^2-1\right)} \; .
\end{eqnarray}
This expression for~$r$ leads to formal fulfilment of the \cterm condition~(\ref{eq:AnniCrea_psi_1})
for all pure squeezed vacuum states, at the price of the outcome state~$\hat a \cdot \psi(x)$ having
the `wrong' orientation for position-squeezed input states.

\subsection{No vacuum state, no displaced states~\label{subsec:NoVacNoDisplaced}}

The case $\sigma_x = 1$ is obviously excluded since that constitutes the vacuum state in
Eq.~(\ref{eq:pure_squeezed_psi}) and whilst in this case the application of~$\hat a^\dag$ generates
the first excited single-photon Fock state~$|1\rangle$, application of $\hat a$ yields zero:
formally, expression~(\ref{eq:AnniCrea_r_condition_1}) diverges.

We notice that displaced states (for which $\langle \hat x \rangle \neq 0$), such as squeezed
coherent states, cannot be solutions of the \cterm condition~(\ref{eq:AnniCrea_psi_2}), because its
left-hand-side term, purely linear in $x$, is incompatible with such displacements of the input states.

\section{Studying Mixed Gaussian States in the Wigner Formulation~\label{sec:Wign-Distr-Formulation}}

Mixed squeezed vacuum states are conveniently stu\-died using Wigner's \ps formulation, since in \ps
such states have real gaussian form.

We will now briefly introduce this formulation and its star-product.

Consider a single-mode operator,~$\hat O$, given in coordinate
representation~$\langle x-y| \hat O | x+y \rangle = O(x-y,x+y)$.
\\
\noindent
To map it to \ps we employ the
Wigner-transform,~${\cal W}[\hat O]$~\cite{Hancock_EJP04,Cohen_LectureNotes18,Zachos_book_21},
\begin{align}\label{eq:WignerWeyl_Trafo}
  {\cal W}[\hat O](x,p) =
  \int_{-\infty}^\infty dy\; O(x-\frac{y}{2},x+\frac{y}{2})\; {\rm e}^{\frac{{\rm i}}{\hbar} p y}\, .
\end{align}

If $\hat O$ is a normalized single-mode density matrix~$\hat \rho$, then the associated
\emph{normalized} distribution in the Wigner repre\-sentation is
$W(x,p) \equiv {\cal W}[\hat \rho]/(2 \pi \hbar)$, which can re\-pre\-sent a mixed state and fulfills
$\int \int dx \; dp \; W = 1$~\cite{Hancock_EJP04,Zachos_book_21}.

\NO{Assuming that the hamiltonian ${\cal W}[ \hat H(\hat x, \hat p) ] = H(x,p)$ is smooth enough, namely,
has a global Taylor expansion, the Wigner transform of the von~Neumann time evolution equation
\begin{equation}\label{eq:W_of_vNeumann}
  {\cal W}\left[ \frac{\partial \hat \rho}{\partial {t}} = \frac{1}{{\rm i}\hbar} [\hat H, \hat
    \rho] \right]
  \end{equation}
  is widely known to have the form of `Moyal's
  bracket',~$\{\!\!\{ {H} , W \}\!\!\}$,~\cite{Groenewold_Phys46,Moyal_MPCPS49,Hancock_EJP04,Zachos_book_21}
\begin{equation}\label{eq:moyal_motion}
  \frac{\partial W}{\partial {t}} = \{\!\!\{ {H} , W \}\!\!\}  
  \; ,
\end{equation}
which is of `Moyal-Sine' form
\begin{align}\label{EqMoyalBraket}
  \{\!\!\{ f, g\}\!\!\} 
     = \frac{2}{\hbar} f(x,p) \sin\!\!\left[\! \frac{\hbar}{2} \!\!\left( 
        \overleftarrow{\frac{\partial}{\partial x}} \overrightarrow{\frac{\partial}{\partial p}}
         - \overleftarrow{\frac{\partial}{\partial p}} \overrightarrow{\frac{\partial}{\partial x}}
  \right)\!\!\right] g(x,p) \; ,
\end{align}
with arrows indicating the `direction' of differentiation:
$f\overrightarrow{\frac{\partial}{\partial x}} g = g\overleftarrow{\frac{\partial}{\partial x}} f =
f \frac{\partial}{\partial x} g$.

Let us use shorthand notations for~$\frac{\partial}{\partial_z} = \partial_z$ and the Poisson
bracket of classical mechanics
$\overleftrightarrow{\partial} = \!\!\left( \overleftarrow{\partial_x} \overrightarrow{\partial_p} -
  \overleftarrow{\partial_p} \overrightarrow{\partial_x} \right)\!\!$ in the Moyal
bracket~$\frac{2}{\hbar}\sin(\frac{\hbar}{2}\overleftrightarrow{\partial})$ of
Eq.~(\ref{EqMoyalBraket}).

It is less well known that Moyal's Sine-bracket~(\ref{EqMoyalBraket}) can be decomposed into
Groenewold~\cite{Groenewold_Phys46,Hancock_EJP04,Zachos_book_21} star-products,~$\star$, given by
$ \{\!\!\{ {H} , W \}\!\!\} = \frac{1}{\rm i \hbar} \left( H \star W - W \star H\right)$ with
}

{ We also need to know 
  how hilbert space operator products
  translate into Wigner's \ps.  Wigner transforming such
  products gives us
  $ {\cal W}[\hat A \cdot \hat B](x,p) = {\cal W}[\hat A](x,p) \star {\cal W}[\hat B](x,p)$, where
  $ \star $ stands for  Groenewold's star product~\cite{Groenewold_Phys46,Hancock_EJP04,Zachos_book_21}.
  Its  explicit form is}
\begin{align}\label{Eq:GroenewoldStar}
  \star &\equiv \exp\left[\frac{i\hbar}{2} 
          \overleftrightarrow {\partial} \right]
          = \sum_{n=0}^{\infty} \frac{(i\hbar \overleftrightarrow {\partial})^n}{2^n n!}  
          \; ,
\end{align}
where the differential operator
$\overleftrightarrow{\partial} = \!\!\left( \overleftarrow{\partial_x} \overrightarrow{\partial_p} -
  \overleftarrow{\partial_p} \overrightarrow{\partial_x} \right)$ is the Poisson bracket of
classical mechanics, namely, overhead arrows indicate the `direction' of differentiation:\\
$f\overrightarrow{\partial_x} g = g\overleftarrow{\partial_x} f =
f(x,p) \left(\frac{\partial}{\partial x} g(x,p)\right)$.

A good mnemonic for~Eq.~(\ref{Eq:GroenewoldStar}) is to think of star-products as the Wigner
transforms of operator compositions, specifically, of the operator multiplication
signs~\cite{Hancock_EJP04}: $ {\cal W}\left[ \cdot \right] = \star $.

Groenewold's star product captures how non-commutativity of hilbert space operators gets mapped into
the quantum \ps of Wigner's formulation, it implements operator non-commutativity in \ps~\cite{Groenewold_Phys46,Hancock_EJP04,Zachos_book_21}.

\section{Photon addition and subtraction in \ps \label{sec:a_adag_ps}}

According to Sect.~\ref{sec:Wign-Distr-Formulation} the photon addition operator applied to the
Wigner distribution 
is~\cite{Ole_23photonAddition}
\begin{subequations}
    \label{eq:addPhotonWoverallEq}
\begin{align}
  \!\!\!\!
  \frac{{\cal W}[\hat a^\dag \! \cdot \! \hat \rho \! \cdot \! \hat a ]}{2 \pi \hbar } = \; & a^*  \star W \star a =  \; \frac{1}{2} \left(p^2+x^2+1\right) W \\ \label{eq:addPhotonW}
& + \left[ -\frac{1}{2} {\VEC \nabla} \cdot \left(\begin{array}{c} x W
                                                       \\
                                                       p W
                                                     \end{array}\right)
   + \frac{\VEC \Delta W}{8} \right]    \; .
  \end{align}
  \end{subequations}
Similarly, the \ps form for photon removal from the light field~$W$ is given by the expression~\cite{Ole_23photonAddition}
\begin{subequations}
    \label{eq:removePhotonWoverallEq}
    \begin{align}\label{eq:a_W_aD_singular}
      \!\!\!\!
      \frac{{\cal W}[\hat a  \! \cdot \! \hat \rho  \! \cdot \! \hat a^\dag ]}{2 \pi \hbar } = \; &
                                                                               a \star W \star a^* =  \; \frac{1}{2} \left(p^2+x^2-1\right) W \\
                                                                             & + \left[ +\frac{1}{2} {\VEC \nabla} \cdot \left(\begin{array}{c} x W
                                                                                                                                 \\
   \label{eq:a_W_aD_divergence}
                                                       p W
                                                     \end{array}\right)
   + \frac{\VEC \Delta W}{8} \right]    \; .
\end{align}
\end{subequations}

\subsection{Impure squeezed vacuum states  with photon added or subtracted\label{subsec:impureSqueeVacAddSub}}

In \ps, the \cterm condition equivalent to Eq.~(\ref{eq:AnniCrea_psi_1}), but now with the
ratio of norms of the form $R = |\!|\hat a^\dag \psi |\!|_2 / |\!|\hat a \psi |\!|_2$, reads
\begin{align}\label{eq:_aWad_adWa}
   a^*  \star  W(x,& \; p) \star a  - R \;  a  \star W(x,p) \star a^*  
            \notag \\
 = ({1 - R}) & \left(\frac{p^2 + x^2 +1 }{2}  
            +\frac{\VEC \Delta }{8}+\frac{p}{2} \frac{\partial
            }{\partial p}+\frac{x}{2}
            \frac{\partial }{\partial x}\right) W(x,p)
            \notag \\
  - & \left( 1 + p \frac{\partial }{\partial p} + x   \frac{\partial }{\partial x} \right) W(x,p) \;  = \;  0 \; .
\end{align}
Using \emph{impure} squeezed vacuum states ($ \sigma_x \sigma_p  > 1$) 
\begin{align}\label{Eq:ImPureSqueezedVacuumState_W}
W( x, p, \sigma_x, \sigma_p) = {\exp[ -\frac{x^2}{{ \sigma_x }^2} -\frac{p^2}{\sigma_p^2} ]}/({\pi {\sigma_p} { \sigma_x
  }}) \; ,
\end{align}
as inputs for Eq.~(\ref{eq:addPhotonWoverallEq}), yields, for \emph{normalized} photon-added outcome
states, the expression
\begin{subequations}\label{Eq:ImPureSqueezedVacuumState_W_Add}
\begin{align}\tag{\ref{Eq:ImPureSqueezedVacuumState_W_Add}}
  W_+( x, p, \sigma_x, \sigma_p) = f_+( x, &p, \sigma_x, \sigma_p)  W( x, p, \sigma_x, \sigma_p) \\
  \text{where }  f_+( x, p, \sigma_x, \sigma_p) = &
\frac{2 p^2 \left({\sigma_p}^2+1\right)^2}{{\sigma_p}^4
  \left({\sigma_p}^2+{\sigma_x}^2+2\right)}
  \\
  -& \frac{{\sigma_p}^2 \left(2 {\sigma_x}^2+1\right)+{\sigma_x}^2}{{\sigma_p}^2 {\sigma_x}^2 \left(
  {\sigma_p}^2+{\sigma_x}^2+2\right)}\\
  +& \frac{2 \left({\sigma_x}^2+1\right)^2 x^2}{{\sigma_x}^4 \left({\sigma_p}^2+
  {\sigma_x}^2+2\right)} \; , 
\end{align}
\end{subequations}
and Eq.~(\ref{eq:removePhotonWoverallEq}), for \emph{normalized} photon-subtracted outcome states,
yields the expression
\begin{subequations}\label{Eq:ImPureSqueezedVacuumState_W_Sub}
\begin{flalign}\tag{\ref{Eq:ImPureSqueezedVacuumState_W_Sub}}
  W_-( x, p, \sigma_x, \sigma_p) = f_-( x, &p, \sigma_x, \sigma_p)  W( x, p, \sigma_x, \sigma_p) \\
  \text{where }  f_-( x, p, \sigma_x, \sigma_p) = &  \frac{2 p^2 \left({\sigma_p}^2-1\right)^2}{{\sigma_p}^4
  \left({\sigma_p}^2+{\sigma_x}^2-2\right)} \\
  + &
  \frac{{\sigma_p}^2 \left(2 {\sigma_x}^2-1\right)-{\sigma_x}^2}{{\sigma_p}^2 {\sigma_x}^2
  \left({\sigma_p}^2+{\sigma_x}^2-2\right)}\\
  + &  \frac{2 \left({\sigma_x}^2-1\right)^2 x^2}{{\sigma_x}^4 \left({\sigma_p}^2+{\sigma_x}^2-2\right)} \; .
\end{flalign}
\end{subequations}
Generally, expressions~(\ref{Eq:ImPureSqueezedVacuumState_W_Add})
and~(\ref{Eq:ImPureSqueezedVacuumState_W_Sub}) are different from each other.

Using the input state~(\ref{Eq:ImPureSqueezedVacuumState_W}) in the \cterm
condition~(\ref{eq:_aWad_adWa}), returns two constraints:
\begin{align}\label{Eq:NormRatio_Wigner}
  R(\sigma_x) = (\sigma_x^2 + 1)^2/(\sigma_x^2 - 1)^2 \; 
\end{align}
and $\sigma_p = 1 / \sigma_x$.

\begin{figure}[t!] \centering
  \includegraphics[width=8.5cm,height=2.5cm]{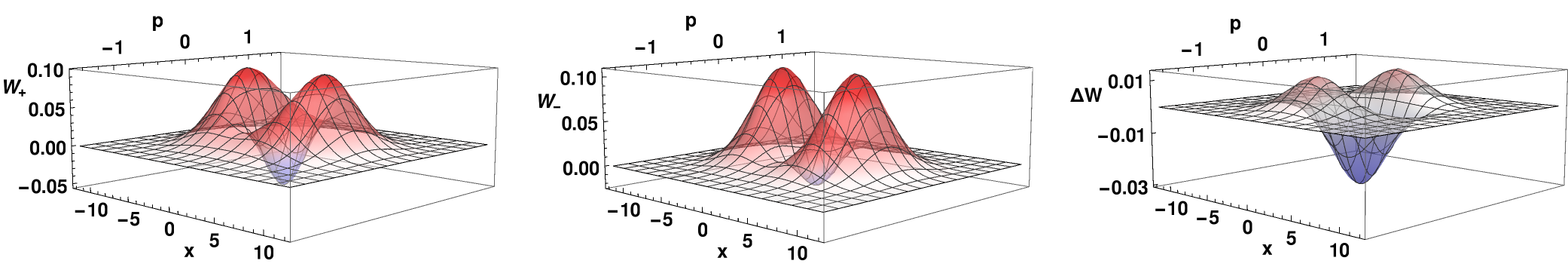}
  \caption{{Using a squeezed vacuum $W(x,p,0,\sigma_x \! = \! 4,\sigma_p \! = \! 1/2)$ input
      state~(\ref{Eq:ImPureSqueezedVacuumState_W}) that is \emph{impure}: Plots of the
      \emph{renormalized} outcome states after a photon is added [$W_+$ of
      Eq.~(\ref{Eq:ImPureSqueezedVacuumState_W_Add}), left panel] or subtracted [$W_-$ of
      Eq.~(\ref{Eq:ImPureSqueezedVacuumState_W_Sub}), middle panel] show that their difference is
      non-zero [$\Delta W = W_+ - W_- \neq 0$, right panel]. The impurity violates the \cterm
      condition~(\ref{eq:_aWad_adWa}).}
   \label{fig:impureSqueezedInput}}
\end{figure}

We notice that the second returned constraint implies that the state is pure:
$\sigma_p \sigma_x = 1$, fulfilling Heisenberg's uncertainty principle~\cite{Lvovsky_16squeezed}
$\langle x^2 \rangle \langle p^2 \rangle = 1/4$, see Fig.~\ref{fig:impureSqueezedInput}.

The normalization constraint, $R$ of (\ref{Eq:NormRatio_Wigner}), contains both branches
$R = (\pm r)^2$ thus `automatically' giving us the full set of solutions: $0< \sigma_x < 1 $ and
$1 < \sigma_x $, without forcing us to have to resolve the sign-problem of
Sect.~\ref{subsec:SignProbelm}.

As an aside, we mention that the normalization constraint
$R(\sigma_x)$ agrees with the general commutator expression
$ \langle \hat a \hat a^\dagger \rangle = \langle \hat a^\dagger \hat a \rangle + 1$.  In \ps
language this can also be shown by subtracting Eq.~(\ref{eq:removePhotonWoverallEq}) from
Eq.~(\ref{eq:addPhotonWoverallEq}) (i.e., setting $R$=1 in Eq.~(\ref{eq:_aWad_adWa})) and
taking into account that the divergence terms vanish at infinity.

Applying the purity constraint to the squeezed vacuum input states [setting $\sigma_p = 1/\sigma_x$
in~(\ref{Eq:ImPureSqueezedVacuumState_W})], we get \cterm, since
$f_+ = f_- = 2 p^2 {\sigma_x}^2 + \frac{2 x^2}{{\sigma_x}^2}-1$ are identical, and so~$W_+=W_-$,
with values $W_{+} = W_{-} = - \frac{1}{\pi}$ at the origin of \ps.
\\
Adding up Wigner distributions of input states (i.e., adding states incoherently) with varying
degrees of squeezing,~$\sigma_x$, violates the \cterm condition~(\ref{eq:_aWad_adWa}), since $f_+$
and $f_-$ depend on~$\sigma_x$ in different ways.

\subsection{Impure gaussian  states \label{subsec:impureGaussian}}

In order to treat a slightly more general mixed-state case we consider the general expression
$W(x,p) = \exp[ g(x,p) ]$.  Since we deal with the second order li\-ne\-ar partial differential
equation~(\ref{eq:_aWad_adWa}), we expect three integration constants plus a fourth constant term in
the exponent, which is fixed by the normalization condition.

We expand $g$ to fourth order in $x$ and $p$ and find that fulfilling
Eq.~(\ref{eq:_aWad_adWa}) implies that only the quadratic terms can be non-zero:
$g = -\frac{1}{2}( x^2 / \sigma_x^2 + c \; x \; p + p^2 / \sigma_p^2)$.

Checking on the conditions for the three integration constants, $\{ \sigma_x^2, c, \sigma_p^2 \}$,
we find, again, that the state has to be pure~$\sigma_p = 1 / \sigma_x$ and that the diagonal term,
$c$, in $g$ has to be a `trigonometric mean'. Therefore, the general family of pure states that obey the
\cterm condition~(\ref{eq:_aWad_adWa}) are randomly oriented pure squeezed vacuum state of the form~
\begin{align}\label{Eq:GeneralPureSqueezedState}
  W(x,p,\theta,\sigma_x,\sigma_p\!\!=\!\!\frac{1}{\sigma_x}) = \frac{1}{\pi} 
  \exp[ -\frac{x'^{\;2}}{ \sigma_x^2} -{ \sigma_x^2 \; p'^{\;2}} ] 
  \; ,
\end{align}
where $x' =x \cos \theta + p \sin \theta $ and  $p' = p \cos \theta - x \sin \theta $ are rotated
coordinates~\cite{Ekert_Knight__AMJP89}.

\section{Photon addition and subtraction in the Fock representation \label{sec:PureFock}}

In the Fock representation, pure squeezed vacuum states, $ \psi(x, \sigma_x)$ of
Eq.~(\ref{eq:pure_squeezed_psi}), with the positive squeezing parameter $1/\sigma_x = e^z$, or
$\frac{1 - \sigma_x^2}{1 + \sigma_x^2} = \tanh z = \tanh(\ln(1/\sigma_x))$, have the form~\cite{Lvovsky_16squeezed}
\begin{equation}\label{eq:nsq4}
  | \psi(\sigma_x) \rangle =\frac 1{\sqrt{\cosh z}}\sum\limits_{m=0}^\infty
  (-\tanh z)^m\frac{\sqrt{(2m)!}}{2^m m!}\ket{2m} .
\end{equation}
Forming the ratio of photon-added and photon-subtracted states yields, after a few steps of calculation,
\begin{equation}\label{eq:ratioFock_add_subtract}
  \frac{ \hat a^\dagger \; | \psi(\sigma_x) \rangle }{ \hat a \; | \psi(\sigma_x) \rangle } = - \tanh z \; .
\end{equation}
This shows that despite the fact that photon numbers are not conserved, the \cterm is guaranteed.
We note, that a similarly surprising effect is predicted to exist for two-mode vacuum squeezed
states, where removal of a photon in one mode adds a photon in the other~\cite{Lu__CPL99};
similarly, photon numbers can change without adding or subtracting a
photon~\cite{Shringarpure_Franson_PRA19}.

In the language of Sect.~\ref{subsec:impureGaussian}, this ratio~(\ref{eq:ratioFock_add_subtract})
is, as expected, since, according to Eq.~(\ref{eq:AnniCrea_r_condition_1}) $ r = -1/\tanh(z)$.

The ratio conditions~(\ref{eq:AnniCrea_r_condition_1}) and (\ref{eq:ratioFock_add_subtract}) also
make plausible that mixed states, even impure squeezed vacuum
states~(\ref{Eq:ImPureSqueezedVacuumState_W}) or sums of pure squeezed vacuum states with different
squeezing ratios, cannot fulfil the \cterm condition~(\ref{eq:_aWad_adWa}), since the simultaneous
presence of different squeezing levels, $\sigma_x$, cannot be accommodated by the single
ratio~(\ref{eq:ratioFock_add_subtract}): to graphically illustrate this finding,
Fig.~\ref{fig:impureSqueezedInput} displays the case of an impure squeezed input
state~(\ref{Eq:ImPureSqueezedVacuumState_W}) with $\sigma_x \sigma_p = 2$.

\begin{figure}[t] \centering
  \includegraphics[width=8.5cm,height=3.5cm]{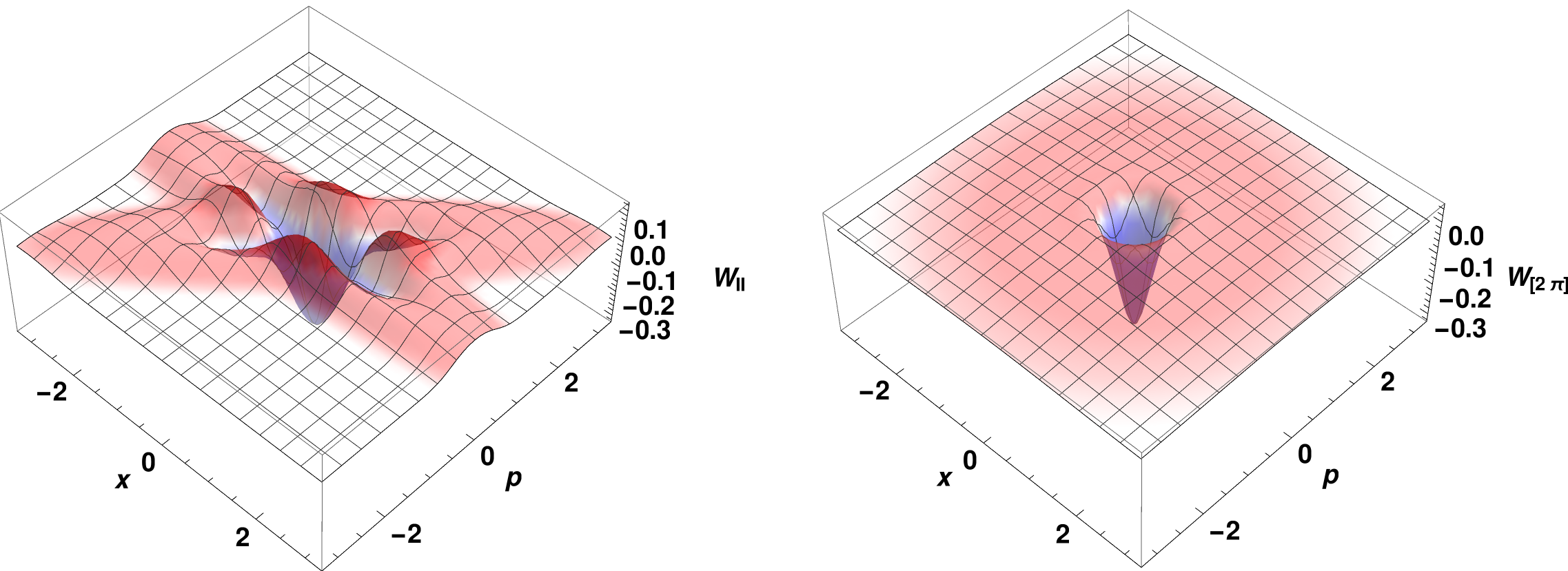}
  \caption{{Plot of photon-added or -subtracted cases for input states $W_{\rm II}$~(\ref{Eq:W_II}),
      with parameters $P=0.5$, $\sigma_x=2.2$, and orienta\-tion angles $\theta_1 = 0$ and
      $\theta_2 = \frac{\pi}{4}$ (left panel) and maximally mixed state
      $W_{[2 \pi ]} (\sigma_x)$~(\ref{Eq:W_2Pi}), with $\sigma_x=2.2$ (right panel). In both cases
      $W[0,0]=-\frac{1}{\pi}$ and both fulfil condition~(\ref{eq:_aWad_adWa}).}
   \label{fig:randomOrientationWsqueezed}}
\end{figure}

\section{Impure Squeezed Vacua\label{sect:ImpureSqueezedVacua}}

The ratio conditions~(\ref{eq:AnniCrea_r_condition_1}), (\ref{eq:ratioFock_add_subtract}),
and~(\ref{Eq:NormRatio_Wigner}) make plausible that only pure squeezed vacuum states are likely
to fulfil the \cterm condition~(\ref{eq:_aWad_adWa}).

Additionally, in Sect.~\ref{subsec:impureGaussian} we showed that expansion to fourth order in the
exponent shows that pure rotated squeezed vacuum states~(\ref{Eq:GeneralPureSqueezedState}) fulfil
the \cterm condition~(\ref{eq:_aWad_adWa}), but, for example, impure squeezed vacuum states of the
gaussian form~(\ref{Eq:ImPureSqueezedVacuumState_W}) do not.

Furthermore, at the end of Sect.~\ref{sec:Schroedinger-Formulation}, we pointed out that pure
displaced states, for which $\langle \hat x \rangle \neq 0$, (or, in light of the rotation
invariance of state~(\ref{Eq:GeneralPureSqueezedState}) for which $\langle \hat p \rangle \neq 0$)
can also not fulfil the \cterm condition~(\ref{eq:_aWad_adWa}).

All this leads us to focus on pure squeezed vacuum states, with the same squeezing, $\sigma_x$, as
candidates for the fulfilment of the \cterm condition~(\ref{eq:_aWad_adWa}).

We notice, that in Wigner \ps the \cterm condition~(\ref{eq:_aWad_adWa}) is a \emph{linear} map of
the input state~$W$.  It can therefore accommodate any two (or more states) simultaneously, i.e., if
they are added up \emph{incoherently}; precisely if those, simultaneously, form the same ratio of
norms~$R$, see Fig.~\ref{fig:randomOrientationWsqueezed}.

The following statement then is the central conclusion of this work:

Any \emph{incoherent} sum of pure squeezed vacuum states~(\ref{Eq:GeneralPureSqueezedState}) with
the same squeezing, $\sigma_x$, but any \ps rotation angle,~$\theta$, can be added up to give impure
states obeying the \cterm condition~(\ref{eq:_aWad_adWa}); pure squeezed vacuum
states~(\ref{Eq:GeneralPureSqueezedState}) are the special case, for pure states.

We furthermore \emph{conjecture} that this covers all possibilities. 

Although we showed that this is true for pure states, we could not prove this conjecture, for mixtures,
because extending the approach of Sect.~\ref{subsec:impureGaussian} by expanding the exponent
function $g$ to fifth or higher order runs into ambiguities, as we explain in the next Section.

\subsection{Mixtures of rotated pure vacuum states with equal squeezing\label{ssec:MixRotatedSqueezedVacua}}

It is straightforward to check explicitly that the line\-a\-rity of the \cterm
condition~(\ref{eq:_aWad_adWa}) admits solu\-tions where the input state is formed from expressions
for pure squeezed states~(\ref{Eq:ImPureSqueezedVacuumState_W_Sub}) of the form~
\begin{align}\label{Eq:W_II}
  W_{\rm II}= P W( \theta_1, \sigma_x) + (1-P) W( \theta_2, \sigma_x) \; ,
\end{align}
with probabilities $0 \leq P \leq
1$ parameterizing the inco\-he\-rent addition, for any orientation
$\theta$ at equal squeezing~$\sigma_x$.

A specific case of $W_{\rm II}$ is displayed in Fig.~\ref{fig:randomOrientationWsqueezed}. Expanding
it two fifth order or higher shows that its exponent function~$g$, see
Sect.~\ref{subsec:impureGaussian}, contains terms of order 5 and above, for both, $x$ and $p$.
Therefore, a proof of our main conjecture just given above, based on the approach of
Sect.~\ref{subsec:impureGaussian} fails. We could not find an argument to exclude the possibility
of other impure states fulfilling the \cterm condition. 

\begin{figure}[h] \centering
  \includegraphics[width=8.5cm,height=3.05cm]{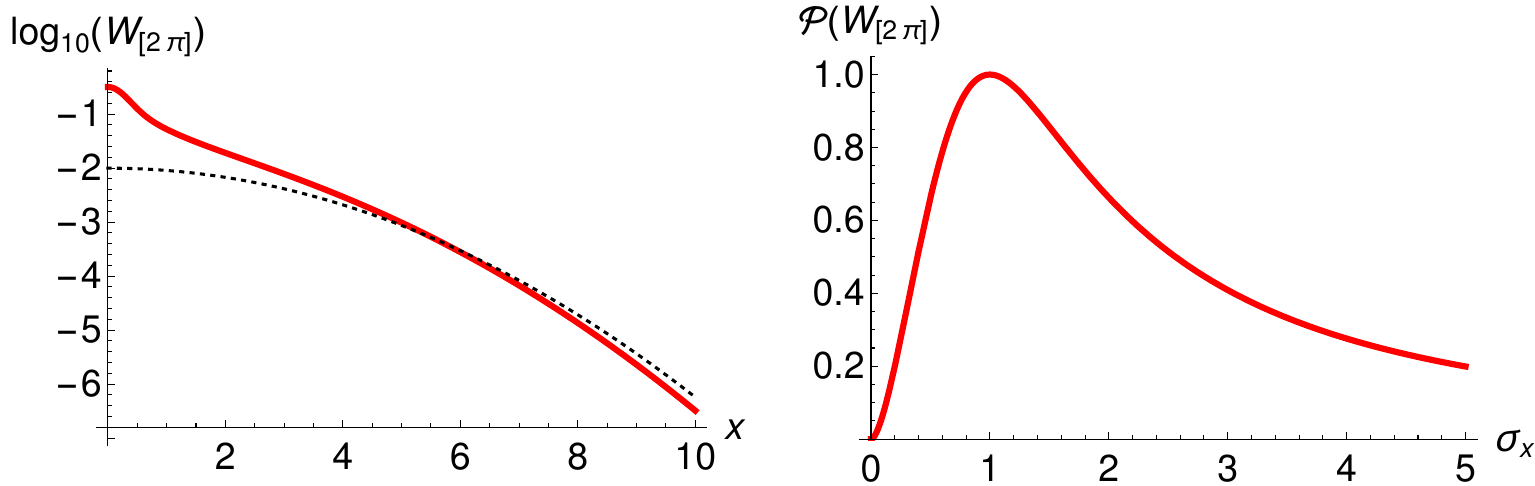}
  \caption{{Radial profile of maximally mixed state~(\ref{Eq:W_2Pi}) log$_{10}[W_{[2 \pi ]} (x, 0, \sigma_x)]$, with
      fixed $\sigma_x=2.2$ (left panel, red curve; for contrast, the black dotted curve shows a
      gaussian profile) demonstrating clear deviation from a gaussian profile. Purity of maximally
      mixed state $\cal P$$(W_{[2 \pi ]} (\sigma_x))$ as a function of squeezing $\sigma_x$ (right panel).}
   \label{fig:MAXrandomOrientationWsqueezed}}
\end{figure}

In this context it is interesting to observe that a simple expression for the maximally-mixed
angular average
$W_{[2\pi]}(\sigma_x) = \frac{1}{2\pi} \int_0^{2\pi} d\theta \; W( x, p,\theta, \sigma_x,
\frac{1}{\sigma_x})$ for pure states~(\ref{Eq:GeneralPureSqueezedState}) with fixed
squeezing~$ \sigma_x$ exists, it is
\begin{align}\label{Eq:W_2Pi}
  W_{[2\pi]} (\sigma_x)  = \; \frac{1}{\pi} & \exp[{-\frac{(x^2+p^2) \left(\sigma_x^4+1\right)}{2 \sigma_x^2}}]
                                              \notag \\
  \times & I_0\left(\frac{(x^2+p^2) \left(\sigma_x^4-1\right)}{2 \sigma_x^2}\right) \; ,
\end{align}
with $I_0$ the
zeroth-order modified Bessel function of the first kind.
\\
Its purity
${\cal P}[W_{[2\pi]}(\sigma_x)] = \int dx \int dp \; 2 \pi W^2_{[2\pi]}(\sigma_x) = (4\sigma_x^2
K[(1 - \sigma_x^4)^2/(1 +\sigma_x^4)^2])/(\pi (1 +\sigma_x^4)) $, where $K$ is the complete elliptic
integral of the first kind, drops with increasing squeezing, see
Fig.~\ref{fig:MAXrandomOrientationWsqueezed}.

Similarly to the case of $W_{\rm II}$, also $W_{[2\pi]} (\sigma_x)$, expanded to high order, contains
terms of 6-th and higher order in $x$ and $p$  in its exponent function~$g$, see
Fig.~\ref{fig:MAXrandomOrientationWsqueezed}.

\section{ Bogoliubov-transformed annihilation operators\label{sect:twoBranches}}

Formally, a squeezed vacuum state~(\ref{eq:pure_squeezed_psi}) can be gene\-ra\-ted by action of the
squeezing operator on the true vacuum $ \hat S(z) | 0 \rangle = | \psi(\sigma_x) \rangle $, with
$z = \ln(1/ \sigma_x)$ real, and 
\begin{equation}\label{SqOp}
 \hat S(\zeta)=\exp[(\zeta\hat a^2-\zeta^*\hat a^{\dag 2})/2] \; . 
\end{equation}
Here $\zeta = z e^{i\phi}$ is the \emph{squeezing parameter}, with
phase~$\phi$~\cite{Lvovsky_16squeezed}. This phase $\phi$ determines the angle of the quadrature that
is being squeezed but for simplicity we can set it to zero for what follows.

It is known that, in the Heisenberg picture, squeezing maps position and momentum according to 
\begin{subequations}\label{QuadTrafoSq}
\begin{eqnarray}
\hat x_z & = \hat S^\dag(z) \; \hat x \; \hat S(z) = \hat x \; e^{-z} & \\ 
\hat p_z & = \hat S^\dag(z) \; \hat p \; \hat S(z) = \hat p \; e^{+z} & \; , 
\end{eqnarray}
\end{subequations}
which describes a passive transformation for the position squeezing by the factor $1/\sigma_x = e^z$,
and corresponding momentum antisqueezing.

The passive squeezing transformation of the creation and annihilation operators is given by
\begin{subequations}
  \label{eq:bogoliubovOperators}\begin{eqnarray}
 \hat a_z &= \hat a \cosh z - \hat a^\dag \sinh z & \\
 \hat a^\dag_z &= \hat a^\dag \cosh z  - \hat a \sinh z &,
\end{eqnarray}\end{subequations}
the well known \emph{Bogoliubov transformations}~\cite{Lvovsky_16squeezed}.

In Sect.~\ref{sec:Schroedinger-Formulation}, we showed that the $\sigma$-parameterized annihilation
operator
\begin{eqnarray}
 \hat a_{\sigma} & = \hat a^\dag - r(\sigma) \; \hat a   \; 
\end{eqnarray}
annihilates pure squeezed vacuum states
of the form~(\ref{eq:pure_squeezed_psi}): $\hat a_{\sigma} \psi(\sigma)=0$ (their  phase is~$\theta=0$).\\
Because of Eq.~(\ref{eq:AnniCrea_r_condition_1}), with $ r = -1/\tanh(z)$, this is of
Bogoliubov-form (where the phase $\phi \neq \theta$ since Eqs.~(\ref{eq:bogoliubovOperators})
describe passive transformations).

A related result was given by Miller and Mishkin in 1966~\cite{Miller_PR66,Dodonov__JOBQSO02}.


What we try to emphasize is this: for the generaliza\-tion to the mixed-state case (here in the
language of \ps), neither the combination $ a_z \star W \star a_z^*$ nor
$ a_\sigma \star W \star a_\sigma^*$ can be used because they contain second order terms,
proportional to $ a^* \star W \star a^*$ and $ a \star W \star a$.

In \ps  the correct form for the \cterm condition is given by equation~(\ref{eq:_aWad_adWa})
which annihilates all $\theta$-rotated pure squeezed vacuum states~(\ref{Eq:GeneralPureSqueezedState})
and mixed states such as~(\ref{Eq:W_II}) and~(\ref{Eq:W_2Pi}).

Therefore, generally speaking, annihilation operators of Bogoliubov form only annihilate properly
aligned pure states, unlike the more general operator $a^* \star W \star a - R(\sigma_x)
 \; a \star W \star a^* $ for the \cterm condition~(\ref{eq:_aWad_adWa}).




\section{Conclusions and Outlook\label{sec:Conclusion}}

Single photon states promise to play a crucial role as photonic qubits~\cite{PhysRevA.102.022620},
as well as providing nonclassical resources for quantum computing~\cite{OBrien-03}. Quantum state
engineering at the level of a few photons also gives rise to the generation of optical cat
states~\cite{kitten, odd, OE-07}.\\
Photon-subtraction is typically realized through a conditional measurement at a beam
splitter~\cite{Dakna__EPJD98, cat-07, lancilla}; but, photon-addition~\cite{Dakna__EPJD98} can also
be realized through the injection of the target states into parametric
down-converters~\cite{Zavatta_SCI04,parigi2007probing,Chen__PRA24,Shringarpure_Franson_PRA19}.

It might appear surprising that adding a photon or subtracting it can give the same outcome,
seemingly viola\-ting energy conservation. However, our calculations using the Fock representation
confirm our results. To consider energy conservation the statistics of an entire ensemble
subjected to subtraction or addition of photons, and its environment, would have to be considered.

Our results do not carry over to further repeated addition or subtraction of photons, since, after the
first round, the resulting state is no longer a pure squeezed state.

Probably there exist other interesting `generalized' symmetries, annihilation operators or
eigen-relations like the \cterm for single-photon addition-versus-subtraction studied here.

\section*{Acknowledgements}

This work is partially supported by the Ministry of Science and Technology of Taiwan (Nos
112-2123-M-007-001, 112-2119-M-008-007, 112-2119-M-007-006), Office of Naval Research Global, the
International Technology Center Indo-Pacific (ITC IPAC) and Army Research Office, under Contract
No. FA5209-21-P-0158, and the collaborative research program of the Institute for Cosmic Ray
Research (ICRR) at the University of Tokyo.

\setcounter{section}{0}
\renewcommand{\thesection}{Appendix~\arabic{section}}
\renewcommand{\thefigure}{A.~\arabic{figure}}
\setcounter{figure}{0}
\setcounter{equation}{0}
\renewcommand{\theequation}{A.\arabic{equation}}










\begin{thebibliography}{25}%
\makeatletter
\providecommand \@ifxundefined [1]{%
 \@ifx{#1\undefined}
}%
\providecommand \@ifnum [1]{%
 \ifnum #1\expandafter \@firstoftwo
 \else \expandafter \@secondoftwo
 \fi
}%
\providecommand \@ifx [1]{%
 \ifx #1\expandafter \@firstoftwo
 \else \expandafter \@secondoftwo
 \fi
}%
\providecommand \natexlab [1]{#1}%
\providecommand \enquote  [1]{``#1''}%
\providecommand \bibnamefont  [1]{#1}%
\providecommand \bibfnamefont [1]{#1}%
\providecommand \citenamefont [1]{#1}%
\providecommand \href@noop [0]{\@secondoftwo}%
\providecommand \href [0]{\begingroup \@sanitize@url \@href}%
\providecommand \@href[1]{\@@startlink{#1}\@@href}%
\providecommand \@@href[1]{\endgroup#1\@@endlink}%
\providecommand \@sanitize@url [0]{\catcode `\\12\catcode `\$12\catcode
  `\&12\catcode `\#12\catcode `\^12\catcode `\_12\catcode `\%12\relax}%
\providecommand \@@startlink[1]{}%
\providecommand \@@endlink[0]{}%
\providecommand \url  [0]{\begingroup\@sanitize@url \@url }%
\providecommand \@url [1]{\endgroup\@href {#1}{\urlprefix }}%
\providecommand \urlprefix  [0]{URL }%
\providecommand \Eprint [0]{\href }%
\providecommand \doibase [0]{https://doi.org/}%
\providecommand \selectlanguage [0]{\@gobble}%
\providecommand \bibinfo  [0]{\@secondoftwo}%
\providecommand \bibfield  [0]{\@secondoftwo}%
\providecommand \translation [1]{[#1]}%
\providecommand \BibitemOpen [0]{}%
\providecommand \bibitemStop [0]{}%
\providecommand \bibitemNoStop [0]{.\EOS\space}%
\providecommand \EOS [0]{\spacefactor3000\relax}%
\providecommand \BibitemShut  [1]{\csname bibitem#1\endcsname}%
\let\auto@bib@innerbib\@empty
\bibitem [{\citenamefont {{Chung}}(2014)}]{Chung_MPLA14}%
  \BibitemOpen
  \bibfield  {author} {\bibinfo {author} {\bibfnamefont {W.~S.}\ \bibnamefont
  {{Chung}}},\ }\bibfield  {title} {\bibinfo {title} {{On the deformed
  photon-added and photon-subtracted states}},\ }\href
  {https://doi.org/10.1142/S0217732314501740} {\bibfield  {journal} {\bibinfo
  {journal} {Mod. Phys. Lett. A}\ }\textbf {\bibinfo {volume} {29}},\ \bibinfo
  {eid} {1450174} (\bibinfo {year} {2014})}\BibitemShut {NoStop}%
\bibitem [{\citenamefont {{Lu}}(1999)}]{Lu__CPL99}%
  \BibitemOpen
  \bibfield  {author} {\bibinfo {author} {\bibfnamefont {H.}~\bibnamefont
  {{Lu}}},\ }\bibfield  {title} {\bibinfo {title} {{Photon Statistics of
  Photon-Added and Photon-Subtracted Two-Mode Squeezed Vacuum State}},\ }\href
  {https://doi.org/10.1088/0256-307X/16/9/009} {\bibfield  {journal} {\bibinfo
  {journal} {Chin. Phys. Lett.}\ }\textbf {\bibinfo {volume} {16}},\ \bibinfo
  {pages} {646} (\bibinfo {year} {1999})}\BibitemShut {NoStop}%
\bibitem [{\citenamefont {Lvovsky}\ \emph {et~al.}(2020)\citenamefont
  {Lvovsky}, \citenamefont {Grangier}, \citenamefont {Ourjoumtsev},
  \citenamefont {Parigi}, \citenamefont {Sasaki},\ and\ \citenamefont
  {Tualle-Brouri}}]{Lvovsky__A20}%
  \BibitemOpen
  \bibfield  {author} {\bibinfo {author} {\bibfnamefont {A.~I.}\ \bibnamefont
  {Lvovsky}}, \bibinfo {author} {\bibfnamefont {P.}~\bibnamefont {Grangier}},
  \bibinfo {author} {\bibfnamefont {A.}~\bibnamefont {Ourjoumtsev}}, \bibinfo
  {author} {\bibfnamefont {V.}~\bibnamefont {Parigi}}, \bibinfo {author}
  {\bibfnamefont {M.}~\bibnamefont {Sasaki}},\ and\ \bibinfo {author}
  {\bibfnamefont {R.}~\bibnamefont {Tualle-Brouri}},\ }\href@noop {} {\bibinfo
  {title} {Production and applications of non-gaussian quantum states of
  light}} (\bibinfo {year} {2020}),\ \Eprint {https://arxiv.org/abs/2006.16985}
  {arXiv:2006.16985} \BibitemShut {NoStop}%
\bibitem [{\citenamefont {Cahill}\ and\ \citenamefont
  {Glauber}(1969)}]{Cahill_PR69a}%
  \BibitemOpen
  \bibfield  {author} {\bibinfo {author} {\bibfnamefont {K.~E.}\ \bibnamefont
  {Cahill}}\ and\ \bibinfo {author} {\bibfnamefont {R.~J.}\ \bibnamefont
  {Glauber}},\ }\bibfield  {title} {\bibinfo {title} {Ordered expansions in
  boson amplitude operators},\ }\href
  {https://doi.org/10.1103/PhysRev.177.1857} {\bibfield  {journal} {\bibinfo
  {journal} {Phys. Rev.}\ }\textbf {\bibinfo {volume} {177}},\ \bibinfo {pages}
  {1857} (\bibinfo {year} {1969})}\BibitemShut {NoStop}%
\bibitem [{\citenamefont {Miller}\ and\ \citenamefont
  {Mishkin}(1966)}]{Miller_PR66}%
  \BibitemOpen
  \bibfield  {author} {\bibinfo {author} {\bibfnamefont {M.~M.}\ \bibnamefont
  {Miller}}\ and\ \bibinfo {author} {\bibfnamefont {E.~A.}\ \bibnamefont
  {Mishkin}},\ }\bibfield  {title} {\bibinfo {title} {Characteristic states of
  the electromagnetic radiation field},\ }\href
  {https://doi.org/10.1103/PhysRev.152.1110} {\bibfield  {journal} {\bibinfo
  {journal} {Phys. Rev.}\ }\textbf {\bibinfo {volume} {152}},\ \bibinfo {pages}
  {1110} (\bibinfo {year} {1966})}\BibitemShut {NoStop}%
\bibitem [{\citenamefont {Hancock}\ \emph {et~al.}(2004)\citenamefont
  {Hancock}, \citenamefont {Walton},\ and\ \citenamefont
  {Wynder}}]{Hancock_EJP04}%
  \BibitemOpen
  \bibfield  {author} {\bibinfo {author} {\bibfnamefont {J.}~\bibnamefont
  {Hancock}}, \bibinfo {author} {\bibfnamefont {M.~A.}\ \bibnamefont
  {Walton}},\ and\ \bibinfo {author} {\bibfnamefont {B.}~\bibnamefont
  {Wynder}},\ }\bibfield  {title} {\bibinfo {title} {{Quantum mechanics another
  way}},\ }\href {https://doi.org/10.1088/0143-0807/25/4/008} {\bibfield
  {journal} {\bibinfo  {journal} {Eur. J. Phys.}\ }\textbf {\bibinfo {volume}
  {25}},\ \bibinfo {pages} {525} (\bibinfo {year} {2004})},\ \Eprint
  {https://arxiv.org/abs/physics/0405029} {physics/0405029} \BibitemShut
  {NoStop}%
\bibitem [{\citenamefont {Cohen}(2018)}]{Cohen_LectureNotes18}%
  \BibitemOpen
  \bibfield  {author} {\bibinfo {author} {\bibfnamefont {D.}~\bibnamefont
  {Cohen}},\ }\href@noop {} {\bibinfo {title} {Lecture notes in quantum
  mechanics}} (\bibinfo {year} {2018}),\ \Eprint
  {https://arxiv.org/abs/quant-ph/0605180} {quant-ph/0605180} \BibitemShut
  {NoStop}%
\bibitem [{\citenamefont {Curtright}\ \emph {et~al.}(2014)\citenamefont
  {Curtright}, \citenamefont {Fairlie},\ and\ \citenamefont
  {Zachos}}]{Zachos_book_21}%
  \BibitemOpen
  \bibfield  {author} {\bibinfo {author} {\bibfnamefont {T.~L.}\ \bibnamefont
  {Curtright}}, \bibinfo {author} {\bibfnamefont {D.~B.}\ \bibnamefont
  {Fairlie}},\ and\ \bibinfo {author} {\bibfnamefont {C.~K.}\ \bibnamefont
  {Zachos}},\ }\href
  {https://www.researchgate.net/publication/260264239_A_concise_treatise_on_quantum_mechanics_in_phase_space}
  {\emph {\bibinfo {title} {A concise Treatise on Quantum Mechanics in Phase
  Space}}}\ (\bibinfo  {publisher} {World Scientific; Singapore},\ \bibinfo
  {year} {2014})\BibitemShut {NoStop}%
\bibitem [{\citenamefont {Groenewold}(1946)}]{Groenewold_Phys46}%
  \BibitemOpen
  \bibfield  {author} {\bibinfo {author} {\bibfnamefont {H.~J.}\ \bibnamefont
  {Groenewold}},\ }\bibfield  {title} {\bibinfo {title} {{On the principles of
  elementary quantum mechanics}},\ }\href
  {https://doi.org/10.1016/S0031-8914(46)80059-4} {\bibfield  {journal}
  {\bibinfo  {journal} {Physica}\ }\textbf {\bibinfo {volume} {12}},\ \bibinfo
  {pages} {405} (\bibinfo {year} {1946})}\BibitemShut {NoStop}%
\bibitem [{\citenamefont {Steuernagel}\ and\ \citenamefont
  {Lee}(2023)}]{Ole_23photonAddition}%
  \BibitemOpen
  \bibfield  {author} {\bibinfo {author} {\bibfnamefont {O.}~\bibnamefont
  {Steuernagel}}\ and\ \bibinfo {author} {\bibfnamefont {R.-K.}\ \bibnamefont
  {Lee}},\ }\href@noop {} {\bibinfo {title} {Photon creation viewed from
  wigner’s phase space current perspective}} (\bibinfo {year} {2023}),\
  \Eprint {https://arxiv.org/abs/quant-ph/2307.16510} {quant-ph/2307.16510}
  \BibitemShut {NoStop}%
\bibitem [{\citenamefont {Lvovsky}(2016)}]{Lvovsky_16squeezed}%
  \BibitemOpen
  \bibfield  {author} {\bibinfo {author} {\bibfnamefont {A.~I.}\ \bibnamefont
  {Lvovsky}},\ }\href@noop {} {\bibinfo {title} {Squeezed light}} (\bibinfo
  {year} {2016}),\ \Eprint {https://arxiv.org/abs/1401.4118} {arXiv:1401.4118
  [quant-ph]} \BibitemShut {NoStop}%
\bibitem [{\citenamefont {{Ekert}}\ and\ \citenamefont
  {{Knight}}(1989)}]{Ekert_Knight__AMJP89}%
  \BibitemOpen
  \bibfield  {author} {\bibinfo {author} {\bibfnamefont {A.~K.}\ \bibnamefont
  {{Ekert}}}\ and\ \bibinfo {author} {\bibfnamefont {P.~L.}\ \bibnamefont
  {{Knight}}},\ }\bibfield  {title} {\bibinfo {title} {{Correlations and
  squeezing of two-mode oscillations}},\ }\href
  {https://doi.org/10.1119/1.15922} {\bibfield  {journal} {\bibinfo  {journal}
  {Am. J. Phys.}\ }\textbf {\bibinfo {volume} {57}},\ \bibinfo {pages} {692}
  (\bibinfo {year} {1989})}\BibitemShut {NoStop}%
\bibitem [{\citenamefont {Shringarpure}\ and\ \citenamefont
  {Franson}(2019)}]{Shringarpure_Franson_PRA19}%
  \BibitemOpen
  \bibfield  {author} {\bibinfo {author} {\bibfnamefont {S.~U.}\ \bibnamefont
  {Shringarpure}}\ and\ \bibinfo {author} {\bibfnamefont {J.~D.}\ \bibnamefont
  {Franson}},\ }\bibfield  {title} {\bibinfo {title} {Generating photon-added
  states without adding a photon},\ }\href
  {https://doi.org/10.1103/PhysRevA.100.043802} {\bibfield  {journal} {\bibinfo
   {journal} {Phys. Rev. A}\ }\textbf {\bibinfo {volume} {100}},\ \bibinfo
  {pages} {043802} (\bibinfo {year} {2019})}\BibitemShut {NoStop}%
\bibitem [{\citenamefont {Dodonov}(2002)}]{Dodonov__JOBQSO02}%
  \BibitemOpen
  \bibfield  {author} {\bibinfo {author} {\bibfnamefont {V.~V.}\ \bibnamefont
  {Dodonov}},\ }\bibfield  {title} {\bibinfo {title} {`nonclassical' states in
  quantum optics: a `squeezed' review of the first 75 years},\ }\bibfield
  {journal} {\bibinfo  {journal} {J. Opt. B: Quant. Semiclass. Opt.}\ }\textbf
  {\bibinfo {volume} {4}},\ \href {https://doi.org/10.1088/1464-4266/4/1/201}
  {10.1088/1464-4266/4/1/201} (\bibinfo {year} {2002})\BibitemShut {NoStop}%
\bibitem [{\citenamefont {Hsu}\ \emph {et~al.}(2020)\citenamefont {Hsu},
  \citenamefont {Lai}, \citenamefont {Chang}, \citenamefont {Wu},\ and\
  \citenamefont {Lee}}]{PhysRevA.102.022620}%
  \BibitemOpen
  \bibfield  {author} {\bibinfo {author} {\bibfnamefont {L.-Y.}\ \bibnamefont
  {Hsu}}, \bibinfo {author} {\bibfnamefont {C.-Y.}\ \bibnamefont {Lai}},
  \bibinfo {author} {\bibfnamefont {Y.-C.}\ \bibnamefont {Chang}}, \bibinfo
  {author} {\bibfnamefont {C.-M.}\ \bibnamefont {Wu}},\ and\ \bibinfo {author}
  {\bibfnamefont {R.-K.}\ \bibnamefont {Lee}},\ }\bibfield  {title} {\bibinfo
  {title} {Carrying an arbitrarily large amount of information using a single
  quantum particle},\ }\href {https://doi.org/10.1103/PhysRevA.102.022620}
  {\bibfield  {journal} {\bibinfo  {journal} {Phys. Rev. A}\ }\textbf {\bibinfo
  {volume} {102}},\ \bibinfo {pages} {022620} (\bibinfo {year}
  {2020})}\BibitemShut {NoStop}%
\bibitem [{\citenamefont {O'Brien}\ \emph {et~al.}(2003)\citenamefont
  {O'Brien}, \citenamefont {Pryde}, \citenamefont {White}, \citenamefont
  {Ralph},\ and\ \citenamefont {Branning}}]{OBrien-03}%
  \BibitemOpen
  \bibfield  {author} {\bibinfo {author} {\bibfnamefont {J.~L.}\ \bibnamefont
  {O'Brien}}, \bibinfo {author} {\bibfnamefont {G.~J.}\ \bibnamefont {Pryde}},
  \bibinfo {author} {\bibfnamefont {A.~G.}\ \bibnamefont {White}}, \bibinfo
  {author} {\bibfnamefont {T.~C.}\ \bibnamefont {Ralph}},\ and\ \bibinfo
  {author} {\bibfnamefont {D.}~\bibnamefont {Branning}},\ }\bibfield  {title}
  {\bibinfo {title} {Demonstration of an all-optical quantum controlled-not
  gate},\ }\href {https://doi.org/10.1038/nature02054} {\bibfield  {journal}
  {\bibinfo  {journal} {Nature}\ }\textbf {\bibinfo {volume} {426}},\ \bibinfo
  {pages} {1476} (\bibinfo {year} {2003})}\BibitemShut {NoStop}%
\bibitem [{\citenamefont {Ourjoumtsev}\ \emph {et~al.}(2006)\citenamefont
  {Ourjoumtsev}, \citenamefont {Tualle-Brouri}, \citenamefont {Laurat},\ and\
  \citenamefont {Grangier}}]{kitten}%
  \BibitemOpen
  \bibfield  {author} {\bibinfo {author} {\bibfnamefont {A.}~\bibnamefont
  {Ourjoumtsev}}, \bibinfo {author} {\bibfnamefont {R.}~\bibnamefont
  {Tualle-Brouri}}, \bibinfo {author} {\bibfnamefont {J.}~\bibnamefont
  {Laurat}},\ and\ \bibinfo {author} {\bibfnamefont {P.}~\bibnamefont
  {Grangier}},\ }\bibfield  {title} {\bibinfo {title} {Generating optical
  schr\"odinger kittens for quantum information processing},\ }\href
  {https://doi.org/10.1126/science.1122858} {\bibfield  {journal} {\bibinfo
  {journal} {Science}\ }\textbf {\bibinfo {volume} {312}},\ \bibinfo {pages}
  {83} (\bibinfo {year} {2006})}\BibitemShut {NoStop}%
\bibitem [{\citenamefont {Neergaard-Nielsen}\ \emph {et~al.}(2006)\citenamefont
  {Neergaard-Nielsen}, \citenamefont {Nielsen}, \citenamefont {Hettich},
  \citenamefont {M{\o}lmer},\ and\ \citenamefont {Polzik}}]{odd}%
  \BibitemOpen
  \bibfield  {author} {\bibinfo {author} {\bibfnamefont {J.~S.}\ \bibnamefont
  {Neergaard-Nielsen}}, \bibinfo {author} {\bibfnamefont {B.~M.}\ \bibnamefont
  {Nielsen}}, \bibinfo {author} {\bibfnamefont {C.}~\bibnamefont {Hettich}},
  \bibinfo {author} {\bibfnamefont {K.}~\bibnamefont {M{\o}lmer}},\ and\
  \bibinfo {author} {\bibfnamefont {E.~S.}\ \bibnamefont {Polzik}},\ }\bibfield
   {title} {\bibinfo {title} {Generation of a superposition of odd photon
  number states for quantum information networks},\ }\href
  {https://doi.org/10.1103/PhysRevLett.97.083604} {\bibfield  {journal}
  {\bibinfo  {journal} {Phys. Rev. Lett.}\ }\textbf {\bibinfo {volume} {97}},\
  \bibinfo {pages} {083604} (\bibinfo {year} {2006})}\BibitemShut {NoStop}%
\bibitem [{\citenamefont {Wakui}\ \emph {et~al.}(2007)\citenamefont {Wakui},
  \citenamefont {Takahashi}, \citenamefont {Furusawa},\ and\ \citenamefont
  {Sasaki}}]{OE-07}%
  \BibitemOpen
  \bibfield  {author} {\bibinfo {author} {\bibfnamefont {K.}~\bibnamefont
  {Wakui}}, \bibinfo {author} {\bibfnamefont {H.}~\bibnamefont {Takahashi}},
  \bibinfo {author} {\bibfnamefont {A.}~\bibnamefont {Furusawa}},\ and\
  \bibinfo {author} {\bibfnamefont {M.}~\bibnamefont {Sasaki}},\ }\bibfield
  {title} {\bibinfo {title} {Photon subtracted squeezed states generated with
  periodically poled \text{KTiOPO}$_4$},\ }\href
  {https://doi.org/10.1364/OE.15.003568} {\bibfield  {journal} {\bibinfo
  {journal} {Opt. Express}\ }\textbf {\bibinfo {volume} {15}},\ \bibinfo
  {pages} {3568} (\bibinfo {year} {2007})}\BibitemShut {NoStop}%
\bibitem [{\citenamefont {Dakna}\ \emph {et~al.}(1998)\citenamefont {Dakna},
  \citenamefont {Knöll},\ and\ \citenamefont {Welsch}}]{Dakna__EPJD98}%
  \BibitemOpen
  \bibfield  {author} {\bibinfo {author} {\bibfnamefont {M.}~\bibnamefont
  {Dakna}}, \bibinfo {author} {\bibfnamefont {L.}~\bibnamefont {Knöll}},\ and\
  \bibinfo {author} {\bibfnamefont {D.-G.}\ \bibnamefont {Welsch}},\ }\bibfield
   {title} {\bibinfo {title} {Quantum state engineering using conditional
  measurement on a beam splitter},\ }\bibfield  {journal} {\bibinfo  {journal}
  {Eur. Phys. J. D - Atom. Mol. Opt. Phys.}\ }\textbf {\bibinfo {volume} {3}},\
  \href {https://doi.org/10.1007/s100530050177} {10.1007/s100530050177}
  (\bibinfo {year} {1998})\BibitemShut {NoStop}%
\bibitem [{\citenamefont {Ourjoumtsev}\ \emph {et~al.}(2007)\citenamefont
  {Ourjoumtsev}, \citenamefont {Jeong}, \citenamefont {Tualle-Brouri},\ and\
  \citenamefont {Grangier}}]{cat-07}%
  \BibitemOpen
  \bibfield  {author} {\bibinfo {author} {\bibfnamefont {A.}~\bibnamefont
  {Ourjoumtsev}}, \bibinfo {author} {\bibfnamefont {H.}~\bibnamefont {Jeong}},
  \bibinfo {author} {\bibfnamefont {R.}~\bibnamefont {Tualle-Brouri}},\ and\
  \bibinfo {author} {\bibfnamefont {P.}~\bibnamefont {Grangier}},\ }\bibfield
  {title} {\bibinfo {title} {Generation of optical `schr\"odinger cats' from
  photon number states},\ }\href {https://doi.org/210.1126/science.1122858}
  {\bibfield  {journal} {\bibinfo  {journal} {Nature}\ }\textbf {\bibinfo
  {volume} {448}},\ \bibinfo {pages} {784} (\bibinfo {year}
  {2007})}\BibitemShut {NoStop}%
\bibitem [{\citenamefont {Takahashi}\ \emph {et~al.}(2008)\citenamefont
  {Takahashi}, \citenamefont {Wakui}, \citenamefont {Suzuki}, \citenamefont
  {Takeoka}, \citenamefont {Hayasaka}, \citenamefont {Furusawa},\ and\
  \citenamefont {Sasaki}}]{lancilla}%
  \BibitemOpen
  \bibfield  {author} {\bibinfo {author} {\bibfnamefont {H.}~\bibnamefont
  {Takahashi}}, \bibinfo {author} {\bibfnamefont {K.}~\bibnamefont {Wakui}},
  \bibinfo {author} {\bibfnamefont {S.}~\bibnamefont {Suzuki}}, \bibinfo
  {author} {\bibfnamefont {M.}~\bibnamefont {Takeoka}}, \bibinfo {author}
  {\bibfnamefont {K.}~\bibnamefont {Hayasaka}}, \bibinfo {author}
  {\bibfnamefont {A.}~\bibnamefont {Furusawa}},\ and\ \bibinfo {author}
  {\bibfnamefont {M.}~\bibnamefont {Sasaki}},\ }\bibfield  {title} {\bibinfo
  {title} {Generation of large-amplitude coherent-state superposition via
  ancilla-assisted photon subtraction},\ }\href
  {https://doi.org/10.1103/PhysRevLett.101.233605} {\bibfield  {journal}
  {\bibinfo  {journal} {Phys. Rev. Lett.}\ }\textbf {\bibinfo {volume} {101}},\
  \bibinfo {pages} {233605} (\bibinfo {year} {2008})}\BibitemShut {NoStop}%
\bibitem [{\citenamefont {Zavatta}\ \emph {et~al.}(2004)\citenamefont
  {Zavatta}, \citenamefont {Viciani},\ and\ \citenamefont
  {Bellini}}]{Zavatta_SCI04}%
  \BibitemOpen
  \bibfield  {author} {\bibinfo {author} {\bibfnamefont {A.}~\bibnamefont
  {Zavatta}}, \bibinfo {author} {\bibfnamefont {S.}~\bibnamefont {Viciani}},\
  and\ \bibinfo {author} {\bibfnamefont {M.}~\bibnamefont {Bellini}},\
  }\bibfield  {title} {\bibinfo {title} {Quantum-to-classical transition with
  single-photon-added coherent states of light},\ }\href
  {https://doi.org/10.1126/science.1103190} {\bibfield  {journal} {\bibinfo
  {journal} {Science}\ }\textbf {\bibinfo {volume} {306}},\ \bibinfo {pages}
  {660} (\bibinfo {year} {2004})}\BibitemShut {NoStop}%
\bibitem [{\citenamefont {Parigi}\ \emph {et~al.}(2007)\citenamefont {Parigi},
  \citenamefont {Zavatta}, \citenamefont {Kim},\ and\ \citenamefont
  {Bellini}}]{parigi2007probing}%
  \BibitemOpen
  \bibfield  {author} {\bibinfo {author} {\bibfnamefont {V.}~\bibnamefont
  {Parigi}}, \bibinfo {author} {\bibfnamefont {A.}~\bibnamefont {Zavatta}},
  \bibinfo {author} {\bibfnamefont {M.}~\bibnamefont {Kim}},\ and\ \bibinfo
  {author} {\bibfnamefont {M.}~\bibnamefont {Bellini}},\ }\bibfield  {title}
  {\bibinfo {title} {Probing quantum commutation rules by addition and
  subtraction of single photons to/from a light field},\ }\href
  {https://doi.org/10.1126/science.1146204} {\bibfield  {journal} {\bibinfo
  {journal} {Science}\ }\textbf {\bibinfo {volume} {317}},\ \bibinfo {pages}
  {1890} (\bibinfo {year} {2007})}\BibitemShut {NoStop}%
\bibitem [{\citenamefont {Chen}\ \emph {et~al.}(2024)\citenamefont {Chen},
  \citenamefont {Hsieh}, \citenamefont {Ning}, \citenamefont {Wu},
  \citenamefont {Chen}, \citenamefont {Shi}, \citenamefont {Yang},
  \citenamefont {Steuernagel}, \citenamefont {Wu},\ and\ \citenamefont
  {Lee}}]{Chen__PRA24}%
  \BibitemOpen
  \bibfield  {author} {\bibinfo {author} {\bibfnamefont {Y.-R.}\ \bibnamefont
  {Chen}}, \bibinfo {author} {\bibfnamefont {H.-Y.}\ \bibnamefont {Hsieh}},
  \bibinfo {author} {\bibfnamefont {J.}~\bibnamefont {Ning}}, \bibinfo {author}
  {\bibfnamefont {H.-C.}\ \bibnamefont {Wu}}, \bibinfo {author} {\bibfnamefont
  {H.~L.}\ \bibnamefont {Chen}}, \bibinfo {author} {\bibfnamefont {Z.-H.}\
  \bibnamefont {Shi}}, \bibinfo {author} {\bibfnamefont {P.}~\bibnamefont
  {Yang}}, \bibinfo {author} {\bibfnamefont {O.}~\bibnamefont {Steuernagel}},
  \bibinfo {author} {\bibfnamefont {C.-M.}\ \bibnamefont {Wu}},\ and\ \bibinfo
  {author} {\bibfnamefont {R.-K.}\ \bibnamefont {Lee}},\ }\bibfield  {title}
  {\bibinfo {title} {Generation of heralded optical cat states by photon
  addition},\ }\href {https://doi.org/10.1103/PhysRevA.110.023703} {\bibfield
  {journal} {\bibinfo  {journal} {Phys. Rev. A}\ }\textbf {\bibinfo {volume}
  {110}},\ \bibinfo {pages} {023703} (\bibinfo {year} {2024})},\ \Eprint
  {https://arxiv.org/abs/quant-ph/2306.13011} {quant-ph/2306.13011}
  \BibitemShut {NoStop}%
\end{thebibliography}
%

\end{document}